\documentclass[usenatbib]{mn2e}
\usepackage{graphicx,times}
\usepackage{amssymb}
\usepackage{natbib}
\usepackage{color}
\usepackage[hidelinks]{hyperref}

\title[Simulating shocks in A1758N]{Simulating the shocks in the dissociative galaxy cluster Abell 1758N}

\author[Machado et al.]{R. E. G. Machado$^{1,2}$\thanks{E-mail: rubens.machado@iag.usp.br}, R. Monteiro-Oliveira$^{1}$, G. B. Lima Neto$^{1}$ and E. S. Cypriano$^{1}$\\$^{1}$Instituto de Astronomia, Geof\'isica e Ci\^encias Atmosf\'ericas, Universidade de S\~ao Paulo, R. do Mat\~ao 1226, 05508-090 S\~ao Paulo, Brazil\\$^{2}$Departamento de Ciencias F\'isicas, Universidad Andr\'es Bello, Av. Rep\'ublica 220, Santiago, Chile
}

\begin{document}  

\date{Accepted 2015 May 19.  Received 2015 May 19; in original form 2015 March 11
}

\pagerange{\pageref{firstpage}--\pageref{lastpage}} \pubyear{2015}

\maketitle

\label{firstpage}

\begin{abstract}
Major mergers between massive clusters have a profound effect in the intracluster gas, which may be used as a probe of the dynamics of structure formation at the high end of the mass function. An example of such a merger is observed at the northern component of Abell 1758, comprised of two massive sub-clusters separated by approximately 750 kpc. One of the clusters exhibits an offset between the dark matter and the intracluster gas.
We aim to determine whether it is possible to reproduce the specific morphological features of this cluster by means of a major merger.
We perform dedicated SPH (smoothed particle hydrodynamics) $N$-body simulations in an attempt to simultaneously recover several observed features of Abell 1758, such as the X-ray morphology and the separation between the two peaks in the projected galaxy luminosity map.
We propose a specific scenario for the off-axis collision of two massive clusters. This model adequately reproduces several observed features and suggests that Abell 1758 is seen approximately 0.4 Gyr after the first pericentric passage, and that the clusters are already approaching their maximum separation. This means that their relative velocity is as low as 380 km s$^{-1}$. At the same time, the simulated model entails shock waves of $\sim$4500 km s$^{-1}$, which are currently undetected presumably due to the low-density medium. We explain the difference between these velocities and argue that the predicted shock fronts, while plausible, cannot be detected from currently available data.

\end{abstract}

\begin{keywords}
methods: numerical -- galaxies: clusters: individual: A1758 -- galaxies: clusters: intracluster medium 
\end{keywords}


\section{Introduction}

In the hierarchical build-up of cosmic structure, cluster mergers are not rare \citep[for a review of cluster formation, see][]{Kravtsov2012}. Much of the observational information about cluster mergers comes from the disturbances in the gas of the intracluster medium (ICM), triggered by collisions \citep[for a review of shocks and cold fronts, see][]{Markevitch2007}. Additionally, gravitational weak lensing analyses (and/or galaxy distribution and dynamics) often supply the crucial input about the total mass distribution, an indispensable ingredient for understanding the dynamics of a cluster merger.

The so-called \textit{dissociative} clusters are the outcome of mergers in which the ICM gas has been spatially separated from the dark matter. While the motion of the collisional gas tends to be hindered by pressure, the collisionless dark matter is able to advance relatively unimpeded. Galaxies are generally expected to accompany the motion of the dark matter. Indeed, even in dissociative cluster mergers, the brightest cluster galaxies (BCGs), or at least the centroids of galaxy light distribution, tend to coincide approximately with mass peaks, unless the morphology is peculiarly complex (for example, in a series of multiple mergers).

There is a growing list of such dissociative clusters in which the gas has been detached from the collisionless components: 1E0657-558 \citep[or Bullet Cluster;][]{Clowe2006}, MACS J0025.4-1222 \citep{Bradac2008}, Abell 520 \citep[or Train Wreck;][]{Mahdavi2007}, Abell 2744 \citep[or Pandora's Cluster;][]{Merten2011}, Abell 2163 \citep{Okabe2011}, Abell 1758N \citep{Ragozzine2012}, DLSCL J0916.2+2951 \citep[or Musket Ball Cluster;][]{Dawson2012} and ACT-CL J0102-4915  \citep[or El Gordo;][]{Jee2014}.

So far, most numerical simulations have focused on the Bullet Cluster itself \citep{Takizawa2005, Takizawa2006, Springel2007, Milosavljevic2007, Mastropietro2008, Akahori2012, Lage2014}. However, the El Gordo cluster has recently been the object of interest in numerical studies \citep{Donnert2014, Molnar2015}. Cluster mergers have often been studied through idealized binary collisions, both to model individual objects, and to investigate general physical processes via parameter space exploration \citep[e.g.][]{ZuHone2010, ZuHone2011}.

An interesting aspect of cluster mergers concerns shock fronts and their velocities. Cluster mergers drive supersonic shock waves of typical Mach numbers $\mathcal{M}\lesssim3$ \citep{Sarazin2002}. Indeed, shocks have been detected in several galaxy clusters using X-ray data from \textit{Chandra} \citep[e.g. Abell 2146;][]{Russell2010}, \textit{Suzaku} \citep[e.g. Abell 3376;][]{Akamatsu2012}, and XMM-\textit{Newton} \citep[e.g. Coma cluster;][]{Ogrean2013}. For these examples, the Mach numbers are in the usual measured range $\mathcal{M} \sim 2$--3, and they are generally estimated from discontinuities in the temperature and/or density profiles, via the Rankine-Hugoniot jump conditions. Projection effects and low photon counts significantly contribute to uncertainties in Mach number measurements. In fact, strong shocks are expected mostly on the periphery of clusters \citep[e.g.][]{Hong2014}, but their detection is made difficult by the low X-ray flux is those regions. 

Radio relics are another observational signature of cluster mergers related to the shock velocity. They are diffuse radio emission, sometimes in the form of arcs, in the periphery of merging clusters and are interpreted as the result of reacceleration of relativistic electrons in the presence of amplified magnetic fields \citep[see][and references therein]{Bruggen2012, Feretti2012}. As pointed out by \cite{Hong2014}, in the cases where shocks are detected both from radio and X-ray data, estimates of shock velocity tend to disagree. Indeed, Mach numbers derived from X-ray data seem to be systematically underestimated with respect to the spectral analysis of radio relics. Recently, \cite{Shimwell2015} have detected a Mpc-scale diffuse radio source on the periphery of the Bullet Cluster. They obtained a Mach number of approximately 5.4 for the main part of this radio relic.

High Mach number shocks are predicted in cosmological simulations \cite[e.g.][]{Vazza2009, Vazza2011}. For example, \cite{Planelles2013} employed shock-capturing techniques on a hydrodynamical cosmological simulation, to study shock statistics. They find the mean mass-weigheted Mach number to be $\mathcal{M}\sim5$ for the entire population of haloes at $z=0$, and values nearly twice as high at $z=1$. Using the recent Illustris simulation suite \citep{Vogelsberger2014,Genel2014}, \cite{Schaal2015} also analysed the statistics of shock waves and found that approximately 75 per cent of shocks have Mach number below  $\mathcal{M}=6$, irrespective of redshift, although the high-$\mathcal{M}$ tail does become more important with increasing redshift. 

The Bullet Cluster itself exhibits a bow shock of $\mathcal{M} \sim 3$, with inferred shock velocity of $\sim$4700 km s$^{-1}$. Taken at face value, this velocity raised concerns that the Bullet Cluster might be a rare outlier, given the small likelihood \citep{Hayashi2006} of such a high-velocity merger for those halo masses in the Lambda Cold Dark Matter ($\Lambda$CDM) cosmology. As \cite{Springel2007} made clear, however, the shock velocity cannot be identified with the velocity of the subcluster. Via numerical simulations, they showed that the speed of the bullet was substantially lower and the reasons were twofold: first, the upstream gas is not at rest; second, the dark matter haloes do not move as fast as the shock front itself. Therefore, they concluded that the actual speed of the bullet must be $\sim$2700 km s$^{-1}$ with respect to the parent cluster. This velocity can be more comfortably accommodated within $\Lambda$CDM expectations, and helped alleviate concerns that new physics would have to be invoked to modify the dark sector. This illustrates the crucial role played by hydrodynamical simulations in understanding cluster mergers and in interpreting the shock velocity.

One of the merging clusters currently known to be dissociative is the northern component of Abell 1758. From \textit{ROSAT} data \citep{Rizza1998}, A1758 was seen to consist of two separate structures: a northern component (A1758N) and a southern component (A1758S). At projected separation of 2 Mpc, these two structures may be gravitationally bound in the strict sense. However, there is no indication at all of interaction between them in the X-ray data \citep{David2004}. Within A1758N itself, a bimodality is seen in the weak lensing mass map, as well as in galaxy number density \citep{Dahle2002}, consistent with galaxy velocities \citep{Boschin2012}. X-ray observations from \textit{Chandra} and XMM-\textit{Newton} \citep{David2004} show that A1758N is undergoing a major merger of two massive clusters, as indicated by gravitational weak lensing results as well \citep{Okabe2008, Ragozzine2012}. This cluster has also been studied from the point of view of metallicity distribution \citep{Durret2011} and galaxy evolution \citep{Haines2009}. Meanwhile, A1758S also displays a disturbed morphology and is undergoing a merger of its own. Finally, A1758N shows diffuse radio emission \citep{Kempner2001, Giovannini2006, Giovannini2009}, which is a sign of a recent merger.

A1758N has been observed intensively during the last 15 years or so, and has had its mass estimated by several authors using different techniques, for instance \citet[][X-rays]{David2004}, \citet[][galaxy member dynamics]{Boschin2012} and \citet[][weak lensing]{Ragozzine2012}. Given that this system is far from hydrostatic equilibrium and that gravitational lensing techniques have difficulty in disentangling the mass of each substructure, there is still a large variance between estimations of each subcomponent of A1758N. It seems safe  to say, however, that the whole A1758N has a mass of the order of $10^{15} M_\odot$ and that there is no huge discrepancy between the masses of its two components.

Our aim in this paper is to model the collision event of A1758N by means of dedicated hydrodynamical $N$-body simulations, attempting to constrain the parameters of the collision (such as initial velocity, impact parameter and time since pericentric passage) and thus to offer one plausible scenario for this merger. More specifically, we wish to determine whether it is possible to reproduce the mophological features of A1758 with a single major merger, or if multiple mergers are necessary.

The simulation setup and initial conditions are described in Section \ref{sec:sim}. Section \ref{sec:results} presents the results concerning global dynamics, X-ray morphology, temperature and shock velocity. In Section \ref{sec:summary} we discuss and summarize our findings. Standard $\Lambda$CDM cosmology (with $\Omega_{\Lambda}=0.7$, $\Omega_{M}=0.3$ and $H_{0}=70$~km~s$^{-1}$~Mpc$^{-1}$) is assumed throughout. The redshift of A1758N is $z=$ 0.279 \citep{Durret2011}, corresponding to an angular scale of 4.233~kpc arcsec$^{-1}$ with our adopted cosmology.

\section{Simulation setup}
\label{sec:sim}

In this paper we focus entirely on the northern structure, A1758N, disregarding the existence of the southern cluster, A1758S. Consequently, we will refer to the two components of A1758N as `the NW cluster' and `the SE cluster'.

To simulate the collision event of A1758N, we set up two spherical galaxy clusters, comprised of dark matter and gas. These are idealized initial conditions, suitable for the systematic study of binary cluster mergers. 


\subsection{Profiles and techniques}

\begin{table}
\caption{Parameters of the cluster initial conditions. Clusters 8 and 9 are the ones used in the best model.}
\label{tb:ic}
\begin{center}
\begin{tabular}{l c c c c c}
\hline
cluster  & $r_{h}$ & $r_{c}$ & $c$ & $M_{200}$            & $r_{200}$ \\ 
         & (kpc)   & (kpc)   &     & ($M_{\odot}$)        & (kpc)     \\
\hline
1        & 455     & 400     & 4.9 & $5.3 \times 10^{14}$ & 1633      \\
2        & 450     & 400     & 5.0 & $5.2 \times 10^{14}$ & 1628      \\
3        & 455     & 200     & 5.0 & $5.4 \times 10^{14}$ & 1649      \\
4        & 455     & 100     & 4.9 & $5.3 \times 10^{14}$ & 1640      \\
5        & 455     & 350     & 4.8 & $5.0 \times 10^{14}$ & 1611      \\
6        & 450     & 170     & 5.0 & $5.3 \times 10^{14}$ & 1646      \\
7        & 480     & 350     & 4.4 & $5.0 \times 10^{14}$ & 1608      \\
8 (NW)   & 500     & 230     & 4.2 & $5.1 \times 10^{14}$ & 1615      \\
9 (SE)   & 440     & 300     & 5.2 & $5.3 \times 10^{14}$ & 1635      \\
\hline
\end{tabular}
\end{center}
\end{table}

\begin{table}
\caption{Initial conditions of the simulations: model name, clusters (from Table \ref{tb:ic}), impact parameter and initial relative velocity. The best model is run 53.}
\label{tb:ic2}
\begin{center}
\begin{tabular}{c c c c | c c c c}
\hline
run & clusters &   $b$   & $|v_{0}|$ & run & clusters &   $b$   & $|v_{0}|$ \\ 
      &          & (kpc) & (km/s)       &       &          &   (kpc) & (km/s)     \\
\hline
1     & 1 + 2    & 50    & 1500 & 30    & 4 + 2    & 150   & 1900 \\
2     & 1 + 2    & 100   & 1500 & 31    & 4 + 2    & 150   & 1800 \\
3     & 1 + 2    & 300   & 1500 & 32    & 4 + 2    & 150   & 1700 \\
4     & 1 + 2    & 500   & 1500 & 33    & 4 + 2    & 150   & 1600 \\
5     & 1 + 2    & 250   & 2000 & 34    & 4 + 2    & 150   & 1500 \\
6     & 1 + 2    & 500   & 2000 & 36    & 6 + 7    & 0     & 2000 \\
7     & 1 + 2    & 750   & 2000 & 37    & 6 + 7    & 250   & 2000 \\
8     & 1 + 2    & 1000  & 2000 & 38    & 6 + 7    & 500   & 2000 \\
9     & 3 + 2    & 250   & 2000 & 39    & 6 + 7    & 1000  & 2000 \\
10    & 3 + 2    & 500   & 2000 & 40    & 6 + 7    & 200   & 2000 \\
11    & 3 + 2    & 750   & 2000 & 41    & 6 + 7    & 200   & 2100 \\
12    & 3 + 2    & 1000  & 2000 & 42    & 6 + 7    & 200   & 2200 \\
13    & 4 + 2    & 250   & 2000 & 43    & 6 + 7    & 200   & 2300 \\
14    & 4 + 2    & 500   & 2000 & 44    & 6 + 7    & 300   & 1800 \\
15    & 4 + 2    & 750   & 2000 & 45    & 6 + 7    & 300   & 2000 \\
16    & 4 + 2    & 1000  & 2000 & 46    & 6 + 7    & 300   & 2200 \\
17    & 5 + 2    & 250   & 2000 & 48    & 8 + 9    & 100   & 1300 \\
18    & 5 + 2    & 500   & 2000 & 49    & 8 + 9    & 100   & 1500 \\
19    & 5 + 2    & 750   & 2000 & 50    & 8 + 9    & 100   & 1700 \\
20    & 5 + 2    & 1000  & 2000 & 51    & 8 + 9    & 100   & 2000 \\
21    & 5 + 2    & 250   & 1750 & 52    & 8 + 9    & 250   & 1300 \\
22    & 5 + 2    & 500   & 1750 & 53    & 8 + 9    & 250   & 1500 \\
23    & 5 + 2    & 750   & 1750 & 54    & 8 + 9    & 250   & 1700 \\
24    & 5 + 2    & 1000  & 1750 & 55    & 8 + 9    & 250   & 2000 \\
25    & 4 + 2    & 250   & 1900 & 56    & 8 + 9    & 400   & 1300 \\
26    & 4 + 2    & 250   & 1800 & 57    & 8 + 9    & 400   & 1500 \\
27    & 5 + 2    & 250   & 1700 & 58    & 8 + 9    & 400   & 1700 \\
28    & 5 + 2    & 250   & 1600 & 59    & 8 + 9    & 400   & 2000 \\
29    & 5 + 2    & 250   & 1500 &       &          &       &      \\
\hline
\end{tabular}
\end{center}
\end{table}

The dark matter haloes follow a \cite{Hernquist1990} profile:
\begin{equation}
\rho_{h}(r) = \frac{M_{h}}{2 \pi} ~ \frac{r_{h}}{r~(r+r_{h})^{3}} \, ,
\end{equation}
where $M_{h}$ is the total dark matter halo mass, and $r_{h}$ is a scale length. This is similar to the NFW profile \citep*{NFW1997}, except for the outer parts. The gas has a \cite{Dehnen1993} density profile:
\begin{equation}
\rho_{g}(r) = \frac{(3-\gamma)~M_{g}}{4\pi} ~ \frac{r_{g}}{r^{\gamma}(r+r_{g})^{4-\gamma}} \, ,
\end{equation}
where $M_{g}$ is the gas mass and $r_{g}$ is a scale length. For a choice of $\gamma=0$ this resembles the $\beta$-model \citep{Cavaliere1976}. Once the gas density is set, the temperature profile is uniquely defined from the assumption of hydrostatic equilibrium.

Equilibrium initial conditions are created according to the methods described in \cite{Machado2013}, which we briefly outline here. To create numerical realizations of this composite system, we uniformly sample the cumulative mass functions of the dark matter and of the gas, and then assign random directions to the position vectors to obtain the Cartesian coordinates. For the velocities of the dark matter particles, rather than relying on the local Maxwellian approximation, we resort to the distribution function in terms of relative energy $f(\mathcal E)$ by solving the \cite{Eddington1916} formula, where the total gravitational potential (dark matter plus gas) must be taken into account. Via von~Neumann rejection, random pairs of $\mathcal E$ and $f$ are drawn and the accepted values provide the velocities $v^2$. Initial conditions created in this manner are in equilibrium by construction, but allowing each cluster to relax in isolation for a few Gyr ensures that eventual numerical transients will have had time to subside.

Simulations were performed with the smoothed particle hydrodynamics (SPH) $N$-body code {\sc Gadget-2} \citep{Springel2005} using a softening length of $\epsilon = 5$~kpc. The dark matter is represented by collisionless $N$-body particles, while the intracluster medium is represented by an ideal gas with adiabatic index $\gamma=5/3$. Galaxies contribute little \citep[a few per cent, e.g.][]{Lagana2008} to the total mass, and their gravitational influence may be disregarded when studying the global morphology and global dynamics of a cluster merger. Therefore, stars are not present in our simulation and obviously neither are feedback and star formation processes. Magnetic fields are also neglected. Cooling is not included as as first approximation, since its time-scale is longer than the collision time-scale. The evolution is followed for 5 Gyr, but the most relevant phases occur within $\sim$1 Gyr at most. Due to this short time span, and to the spatial scale of the order of only a few Mpc, cosmological expansion is also ignored.


\subsection{Initial conditions for A1758N}

Assuming a total mass of the order of $10^{15} M_{\odot}$ for the A1758N system, we choose to simulate the simplest case of a major merger, i.e. having two equal-mass clusters of approximately $5 \times 10^{14} M_{\odot}$ each. However, for reasons discussed in Section \ref{sec:xray} (namely, the necessity to introduce asymmetry), we must create initial conditions that are somewhat dissimilar. 

We have performed a suite of simulations, briefly summarized here. Table \ref{tb:ic} gives the initial conditions of a set of clusters having roughly the same mass, but different scale lengths of the gas density profiles (and thus different initial gas central densities). Table \ref{tb:ic2} presents the simulation runs in which these clusters were collided with different impact parameters $b$ and initial relative velocities $v_{0}$. These simulations were run with the explicit purpose of obtaining one model whose gas morphology satisfactorily resembles that of A1758N. By the trial-and-error nature of the search, this is not meant to be an exhaustive parameter space exploration. Nevertheless we attempted to cover a sufficient range of physically plausible collision parameters ($0<b<1000$ kpc, $1300<|v_{0}|<2300$ km~s$^{-1}$) while, at the same time, limiting the search to the regimes more likely to produce the target morphology. The simulation outputs were visually inspected and the non-preferred models were discarded mainly on the basis of their inadequate X-ray emission morphology at the required times (see Section \ref{sec:xray}). Models were further refined by requiring a quantitative match of the final gas densities. We found run 53 (using clusters 8 and 9) to be the best model, and in the rest of the paper we focus the discussion almost exclusively on that model.

\begin{figure*}
\includegraphics[height=12cm]{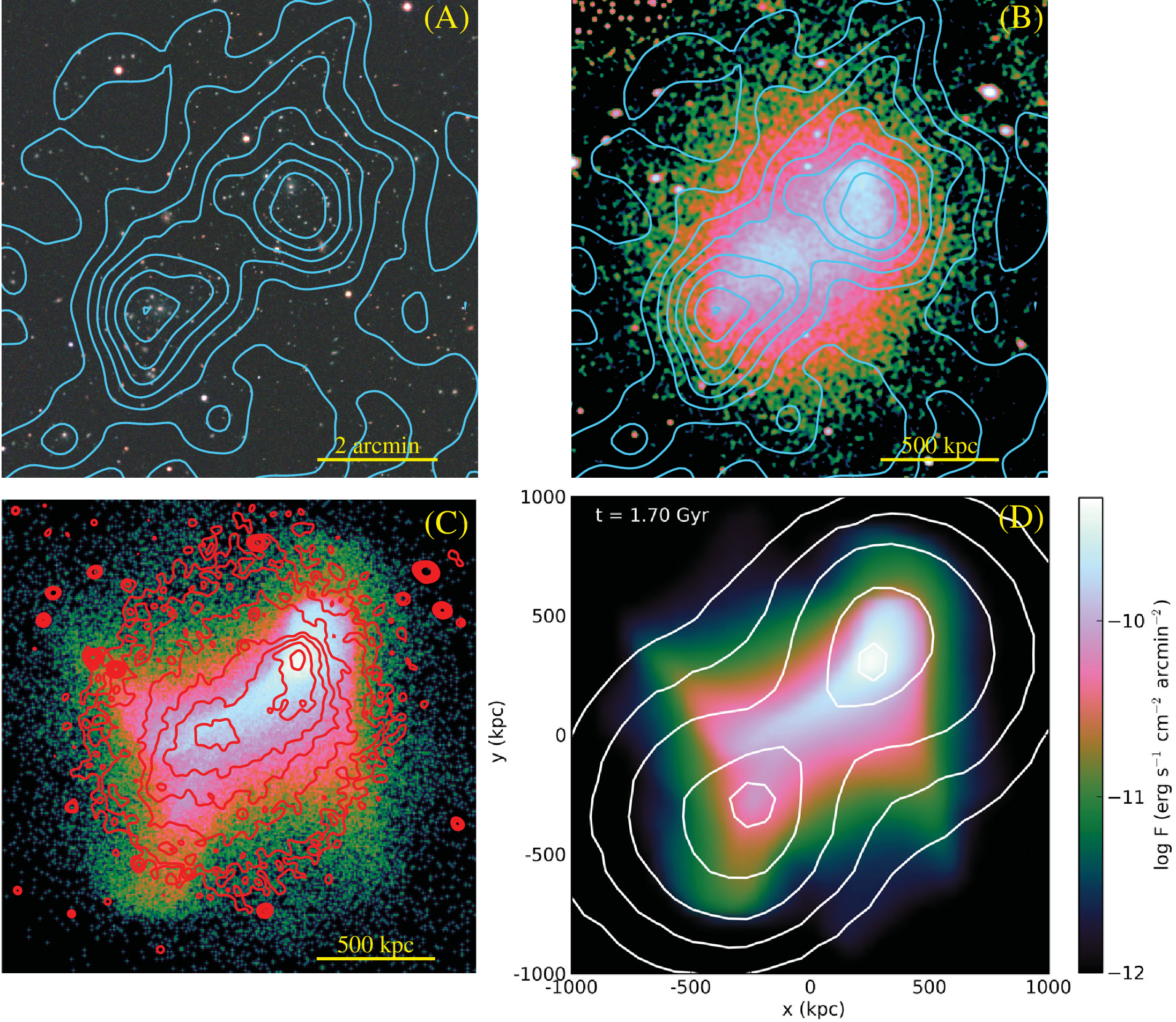}
\caption[]{(a) Optical $g$-$r$-$i$ ``true color'' SDSS image with cyan contours of galaxy luminosity density (isopleths). (b) \textit{Chandra} exposure-map corrected, 0.3--7.0 band X-ray image with contours of galaxy luminosity density. (c) Mock X-ray map created from the simulation and contours of observed X-rays (red lines) based on the \textit{Chandra} image. (d) Simulated X-ray surface brightness (colors) and simulated contours of total mass (white lines). All panels have the same physical size assuming a scale of 4.233~kpc arcsec$^{-1}$.}
\label{fig1}
\end{figure*}

The parameters for the initial conditions of clusters 8 and 9 are given in Table \ref{tb:ic}. For those choices of parameters, the central gas density in cluster 8 is slightly higher than in cluster 9. Their respective central electron number densities are initially 0.05 cm$^{-3}$ and 0.02 cm$^{-3}$. Furthermore, a margin of choice for the concentrations can also be afforded. For the combination of parameters in Table \ref{tb:ic}, the $r_{h}/r_{200}$ ratio is such that the concentrations are consistent with the range $0.6 \lesssim \log c \lesssim 0.7$ for haloes with masses of $\log \left( M_{200} / M_{\odot} \right) \sim 14.7$, considering the redshift $z=0$ \citep{Ludlow2014}.

To prepare the initial configuration ($t=0$), the two clusters are set 3 Mpc apart along the $x$ axis, with an initial impact parameter of $b = 250$ kpc along the $y$ axis. Notice that this impact parameter refers to the initial setup, and the minimum separation at the instant of pericentric passage is considerably smaller. The initial relative velocity is $v_{0} = -1500$ km s$^{-1}$ parallel to the $x$ direction. This is somewhat lower than the velocity of the zero-energy orbit (approximately $-1700$ km s$^{-1}$), i.e. the relative velocity that two point particles of equivalent mass would have at 3 Mpc separation, had they been released from infinity.

The baryon fraction is roughly constant at large radii for both clusters ($0.14-0.16$). Each cluster is represented by $N = 2 \times 10^{6}$ particles, equally divided between dark matter and gas particles. As a convergence test, the model discussed in this paper was re-run with a total of $4 \times 10^{7}$ particles. For the purposes of our analyses, differences in the outcome were not significant.


\section{Results}
\label{sec:results}

We performed a suite of binary merger simulations, and in this paper we report on the particular model that best matched the observations. A number of observed features are successfully reproduced in this model, namely: the current separation between the clusters, the overall X-ray morphology, central gas density and central gas temperatures. 

In order to constrain the simulated system of colliding clusters, we have computed a projected luminosity density map, following the procedure described in \cite{Machado2015}. We have downloaded a table from SDSS-DR10 \footnote{\href{http://skyserver.sdss.org/dr10/en/tools/search/sql.aspx}{http://skyserver.sdss.org/dr10/en/tools/search/sql.aspx}} of all galaxies within $0.2^\circ$ from the mid-point between the two sub-clusters comprising A1758N with $r^\prime$ magnitude between 17.3 (the magnitude of the cluster brightest galaxy) and 23.0. In order to increase the cluster contrast with respect to back and foreground galaxies, we used only objects with colour $0.3 < r-i < 0.7$. Since we were not interesting in a rigorous cluster/field separation we have determined colour range simply by visual inspection of the red-sequence in the $r \times (r-i)$ colour-magnitude diagram. We then strongly smoothed the selected galaxies with a 2D-gaussian kernel with intensity proportional to the galaxy and width proportional to the effective radius (the radius containing half the luminosity of the galaxy). Figure \ref{fig1}a shows the contours of equal projected luminosity over the SDSS $g$-$r$-$i$ image. In this paper, these contours are meant to be understood merely as a proxy for the centroids of cluster mass. The assumption that the galaxies follow the dark matter is commonly verified in merging clusters.


\begin{figure}
\includegraphics[width=\columnwidth]{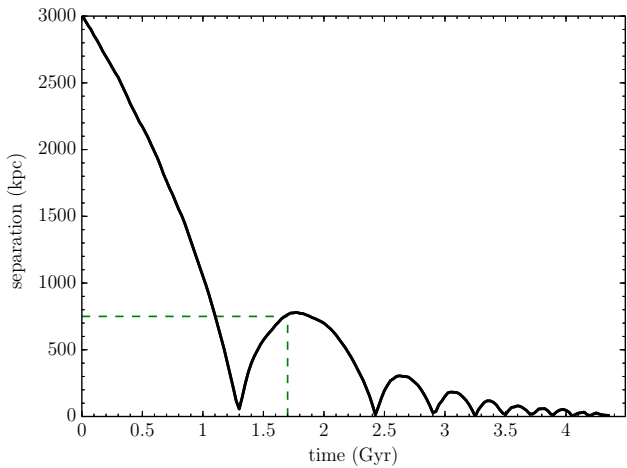}
\caption[]{Separation between the centres of density of the dark matter haloes as a function of time. The first pericentric passage takes place at $t = 1.3$ Gyr and the separation of 750 kpc is reached at $t = 1.7$ Gyr.}
\label{fig2}
\end{figure}

\begin{figure}
\includegraphics[width=\columnwidth]{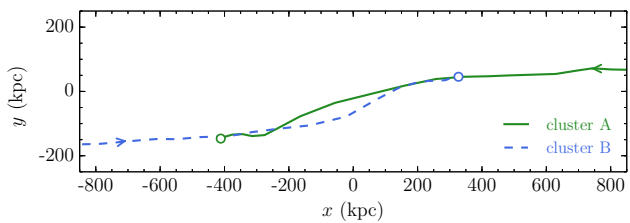}
\caption[]{Orbits of the two clusters in the centre-of-mass rest-frame, as traced by the positions of the dark matter peaks, disregarding the gas. The open circles correspond to the final state at $t = 1.7$ Gyr.}
\label{fig3}
\end{figure}

\subsection{Global dynamics}
\label{sec:global}

From the galaxy luminosity map, we see that the NW and the SE mass peaks are separated by approximately 750 kpc. This value also agrees with the separation of the mass peaks derived from lensing analysis \citep{Okabe2008, Ragozzine2012}, confirming that the galaxy distribution is a good indicator of the mass distribution in the case of A1758N. From the dynamical point of view, this is the major constraint on the simulation model.

As a proxy for the mass centroids of the simulated clusters, we use the density peaks, i.e. the location of the center of density of each cluster's dark matter particles. With this definition, we may straighforwardly measure the separation between the two clusters as a function of time, seen in Fig.~\ref{fig2}. The first pericentric passage takes place at $t = 1.3$ Gyr and the desired separation of 750 kpc is then attained at $t = 1.7$ Gyr. Thus we have the best fit instant occurring 0.4 Gyr after pericentric passage. Strictly speaking, there are also two other instants when the centroids are 750 kpc apart. One of them is at $t = 1.1$ Gyr, prior to the first pericentric passage. At this time, the clusters are still approaching for the first time and no relevant asymmetry has been induced in the morphology yet. Therefore the current stage of the A1758N merger is likely to be post-pericentric passage. The other possible instant would be at $t = 1.9$ Gyr, slightly after the point of return. This is also ruled out because by then the desired morphology is not well reproduced.

By the time the best-fitting separation of 750 kpc is arrived at, the relative velocity between the dark matter haloes has decreased to approximately 380 km s$^{-1}$. Indeed, Fig.~\ref{fig2} indicates that this instant corresponds almost to a point of return. Within a further 0.1 Gyr into the future, the apocenter of 780 kpc would have been reached. The relative velocity between the dark matter haloes must not be confused with the \textit{shock} velocity, which is discussed in detail in Section \ref{sec:shock}. Thus we find that the current separation between the clusters is very nearly the maximum separation, meaning they are almost coming to a halt at this point.

Although the impact parameter in the initial conditions was $b = 200$ kpc, the first core passage at $t = 1.3$ Gyr takes place with a pericentric distance of only about 60 kpc. This may seem an exceedingly small distance, but even small pericentric separations are sufficient to induce substantial asymmetry in the gas \cite[e.g.][]{Machado2015}. The future evolution seen in Fig.~\ref{fig2} indicates that after a few more core passages the fate of these clusters is to merge into one massive system. Within the next 2 Gyr, the apocenters will no longer exceed 100 kpc.

The orbit of the dark matter centroids can be seen in Fig.~\ref{fig3}, where one notices that the deflection angle is far smaller than $90^{\circ}$. At the instant $t = 1.7$ Gyr, the line connecting the centroids makes an angle of only $15^{\circ}$ with respect to the initial collision direction.


\subsection{X-ray morphology}
\label{sec:xray}

One of the crucial observed features of A1758N is the offset between dark matter and gas; more specifically, the offset between the SE cluster and part of the gas that trails behind it. A successful model must then \textit{simultaneously} reproduce the overall gas morphology, the $\sim 750$~kpc separation, as well as the correct gas densities and temperatures. All of this within the constraints of the total mass, which in turn limits the possibilities of gas content. Independently, each of the observed requirements would not be excessively difficult to satisfy. The difficulty lies in finding a combination of initial parameters that reconciles them all in one model at the same instant. The X-ray morphology is the most challenging aspect to reproduce with simulations, because it is quite sensitive to the various collision parameters, but mostly because it is highly time-dependent.

For example, with physically well-motivated initial conditions, it is relatively simple to find a model in which the correct separation will be eventually achieved at some point in time. This could even be forecast to a good approximation by merely mechanical arguments. On the other hand, it is not at all obvious how to anticipate -- even roughly -- the specific shapes of the resulting disturbances in the gas. The much more complex hydrodynamical interactions do not allow for such predictions without recourse to the actual simulations themselves, which are of course computationally demanding.

Bearing in mind these difficulties, we are able to find one model which approximately matches the broad morphological features of the gas at the correct time (i.e. at the correct separation). Although the simulations are adiabatic, in order to generate the visualizations (maps of projected X-ray surface brightness), we assume that the X-ray radiative losses are described by a cooling function, besides the bremsstrahlung emission \citep[for details, see][]{Machado2013}.

The resulting map at $t = 1.7$ Gyr is shown in Fig.~\ref{fig1}d, where the colours represent simulated X-ray surface brightness and the white contour lines represent total projected mass. This should be compared to Fig.~\ref{fig1}b, which displays the observational data: X-rays from \textit{Chandra} and contours of galaxy luminosity density from SDSS.

We have downloaded the publicly available X-ray observations made with \textit{Chandra} and combined them using the task \texttt{merge\_obs} from the Chandra X-ray Center (CXC)\footnote{See \href{http://cxc.harvard.edu/ciao/ahelp/merge\_obs.html}{http://cxc.harvard.edu/ciao/ahelp/merge\_obs.html}},
that merges multiple exposure-corrected images. We have used 5 exposures with ObsId. 2213, 13997, 15538 and 15540 (PI. L. David), and 7710 (PI. G. Garmire), all of them observed in very faint mode with ACIS-I (except exposure 2213, that was made with ACIS-S. The effective exposure time of the final merged dataset was 213~ks. We have produced three images corresponding to soft (0.3--1.0 keV), hard (2.0--7.0 keV) and broad bands (0.3--7.0 keV). Notice that we did not use the task default values for the band energy limits. The broad band image is shown in Fig.~\ref{fig1}b.

The observational data of Fig.~\ref{fig1}b shows that the NW cluster coincides with a peak of X-ray emission, while the SE cluster does not. This is one of the main aspects that the simulated map in Fig. ~\ref{fig1}d is meant to reproduce. Note that here, the galaxy luminosity isopleths (cyan lines from Fig.~\ref{fig1}a) are acting as rough proxies for the mass centroids.

\begin{figure}
\includegraphics[width=\columnwidth]{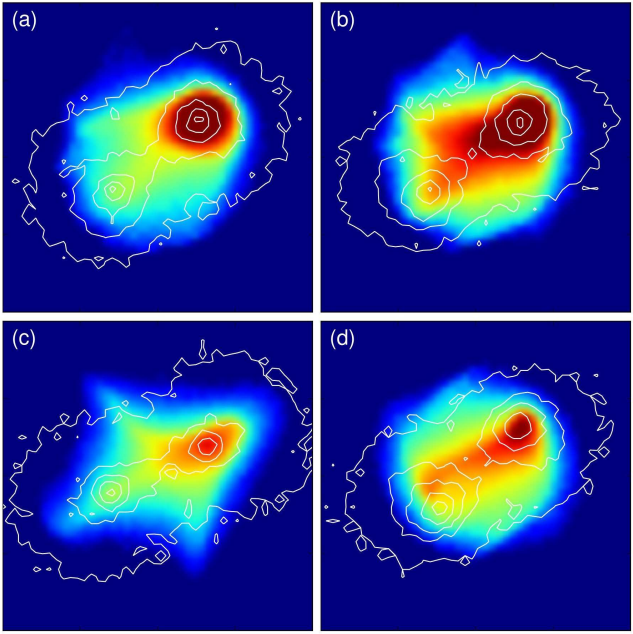}
\caption[]{A sample of non-preferred models at the times when the mass centroid separation is approximately 750 kpc: (a) run 29, (b) run 37, (c) run 48, and (d) run 59. See Table \ref{tb:ic}. Colors represent simulated X-ray surface brightness and white countour lines are the total projected mass. Each panel is 2 Mpc wide.}
\label{fig4}
\end{figure}

To obtain a more realistic comparison between simulation output and observational data, we produced a mock X-ray map using the \texttt{photon\_simulator} module \citep{ZuHone2014} of the \texttt{yt} analysis package\footnote{\href{http://yt-project.org/}{http://yt-project.org/}} \citep{Turk2011}. The procedure, based on \cite{Biffi2012,Biffi2013}, is briefly summarized as follows. Taking as input the simulated gas properties (temperature, density, and also an assumed constant metallicity of 0.3 Z$_{\odot}$), a photon sample is generated assuming a spectral model (the {\sc apec} model from the AtomDB database\footnote{\href{http://www.atomdb.org/}{http://www.atomdb.org/}}). Given the cluster's redshift, the photon sample is projected along a line of sight (in this case, the $z$ axis of the simulated volume). Besides the cosmological redshift, the photon energies are also Doppler-shifted according to the line-of-sight velocities of the gas. An energy-dependent Galactic absorption model is applied, given the coordinates of the cluster. The projected photons are convolved with \textit{Chandra}'s response to create realistic photon counts. We use the same effective exposure time of 213~ks and the same energy range of 0.3--7.0 keV to produce a comparable mock observation shown in  Fig.~\ref{fig1}c. Its morphological features are quite similiar to the simulated X-ray surface brightness map of Fig.~\ref{fig1}d, which was produced assuming essentially bremsstrahlung emission. This similariry indicates that both procedures offer sufficient means of comparison between simulation and observation, at least for our purposes of a broad morphological comparison. However, the mock observation provides a more quantitative picture of where the photon counts are expected to be very low. In Fig.~\ref{fig1}c we also overlay the X-ray surface brightness isocontours from the \textit{Chandra} broad-band image. This helps highlight the shape of the matching regions of highest X-ray emission.

The most striking agreement is the one of the mass countous, which are indeed the most robust result. In the case of the gas morphology, the global shape is fairly well reproduced, with the main peak coinciding with the NW dark matter halo, and with some considerable amount of gas to be found in the region between the two dark matter peaks. Although the general curved shape of that gas is adequate, in both simulation visualizations the secondary X-ray peak seems somewhat less pronounced than in the observation. Quantitatively, we find gas densities in the regions of the X-ray peaks in very acceptable agreement with the values obtained by \cite{David2004}. In terms of electron number density at $t = 1.7$ Gyr we measure (in the NW and SE clusters respectively) 0.017 cm$^{-3}$ and 0.008 cm$^{-3}$ from our simulation, while \cite{David2004} had measured 0.017 cm$^{-3}$ and 0.012 cm$^{-3}$ from the observed clusters. The final morphology is quite sensitive to the initial gas concentrations. Although we employed two clusters with similar virial masses, we found that it was difficult to produce good results unless the central gas densities differed. In order to obtain the necessary asymmetry in the final state, the gas concentration had to be as dissimiar as possible, within the range allowed by observational constraints.

In Fig.~\ref{fig1}, and in the other visualizations, the snapshot at $t = 1.7$ Gyr was rotated $215^{\circ}$ counterclockwise with respect to the original $(x,y)$ frame. As a result, cluster 8 ends up in the NW quadrant, approximately matching the position angle of the observations. This is merely a rotation on the plane of the sky, and the line of sight remains perpendicular to the plane of the orbit. We have found no meaningful improvement in the morphology of the gas by projecting the system under an inclination angle and therefore the $i = 0^{\circ}$ case remains the preferred solution. Nevertheless, this model would also be consistent with small inclination angles $i \lesssim 20^{\circ}$, albeit with a slightly inferior quality in the morphological match. If that were the case, out of the 380 km s$^{-1}$ relative 3D velocity, as much as $\sim$130 km s$^{-1}$ could be projected along the line of sight.

Our central goal was to determine whether the collision event of A1758N could be reasonably modelled by one single major merger. In search of a good match, we performed numerous simulations, discarding inadequate models. The criterion to reject models relied mainly on visual inspection of the morphology of the X-ray surface brightness maps. We cannot offer confidence intervals of the parameters in a rigorous statistical sense, but Table \ref{tb:ic2} gives an approximate idea of the typical ranges that were explored and rejected. As an illustrative example, we present a small sample of non-preferred models in Fig.~\ref{fig4}. They are shown at the times when the mass centroid separation is approximately 750 kpc in each case. Since this is the strongest dynamical constraint, most models were rejected because their X-ray morphology was inadequate at those instants. In many cases, the SE cluster was excessively stripped during the pericentric passage. For example, Fig.~\ref{fig4}a shows a model with the same impact parameter and the same initial velocity as the best model, but using cluster initial conditions with different central densities. In Fig.~\ref{fig4}b, the collison has an even more concentrated NW cluster, but higher velocity. The models of Figs.~\ref{fig4}c and ~\ref{fig4}d use the same clusters as the best model. But colliding those clusters with smaller impact parameter and smaller velocity (\ref{fig4}c) results in a nearly head-on encounter, while larger impact parameter and larger velocity (\ref{fig4}d) induced excessive asymmetry. In some cases, models with acceptable morphology were discarded on the basis of the quantitative gas density. 

Incidentally, we mention that the simulation result displays a hint of what may be regarded as a ``twin-tailed'' wake. Such tail is a peculiarity of the El Gordo cluster, sought to be reproduced in simulations by \cite{Molnar2015} using an adaptive mesh code. In those simulations, one part of the wake was understood to be tidally stretched gas, and the other, a projection of the compressed gas front. Evidently it was not our aim to obtain this detailed feature, but it is interesting to notice that this emerges in major merger simulations, under certain circumstances. 


\subsection{Temperature}

\begin{figure}
\centering
\includegraphics[width=0.8\columnwidth]{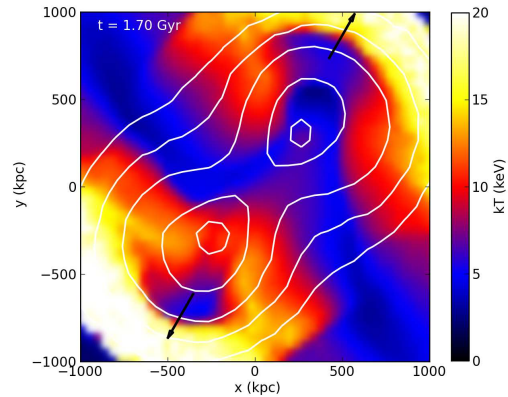}
\caption[]{Emission-weigthed temperature map with contours of total mass. The arrows mark the directions along which the pressure profiles of Fig.~\ref{fig7} were measured.}
\label{fig5}
\end{figure}

From the simulation output, we can evaluate certain quantities that are directly comparable to available observational results, but we may also inspect simulated properties which may be currently undetectable in the observed data. If we measure the average temperatures only in the regions corresponding to the peaks of X-ray emission, we obtain central values of roughly $T \sim 8$ keV for both clusters. This is in sufficient agreement with the results of \cite{David2004}, and is consistent with what one would expect for haloes of such virial masses, also bearing in mind that the gas densities in the same regions were already found to be very well matched. However, from the simulated temperature map, we notice a striking feature having no observational counterpart: two prominent bow shocks, where the shocked gas has been heated to very high temperatures. This is seen in Fig.~\ref{fig5}. The shock fronts are not contiguous with the edges of the X-ray peaks. Rather, each shock front is ahead of the corresponding dark matter halo by a large distance. The outer faces of the shocked gas are to be found almost 1~Mpc ahead of the corresponding dark matter centroids. This means that they are located in regions of very low gas density and are advancing into the coldest parts of the ICM. 

Such high temperatures and strong shocks are generally not observed, although they are not an unusual outcome in simulated cluster mergers. In A1758N no such shocks are detected. We explored several different combinations of initial velocities, impact parameters, etc, within physically plausible ranges. Given the virial masses and gas content of these clusters, we were unable to find models in which strong shocks were not present. Surely one cannot offer proof of the uniqueness of the solution in this type of problem. Nevertheless, a compelling case may be made in favor of this scenario -- or ones broadly similar to it -- given the features that it adequately reproduces, and given the plausibility of the orders of magnitude involved. Yet it seems difficult to escape the rise of strong bow shocks entailed by this kind of collision. If this model can be understood as a reasonable approximation of the circumstances of A1758N, then one must attempt to reconcile its predictions with the absence of detected shocks. The alternative approach would require somehow obtaining a model in which two clusters of $\sim 5 \times 10^{14} M_{\odot}$ and $\sim$8 keV interpenetrate without giving rise to shock-heated gas. We favour the picture in which strong shocks did indeed take place, but have already expanded to the periphery of the cluster, where they may elude detection. Note again that photon counts are nearly non-existent in the outskirts of both the mock X-ray image (Fig.~\ref{fig1}c) and the Chandra observation (Fig.~\ref{fig1}b).

Using \textit{Chandra} data described above (Sec.~\ref{sec:xray}), we have used the soft and hard band images to produce the observed hardness ratio (HR) image:
$$ \mbox{HR} = \left(\mbox{hard} - \mbox{soft}\right) /  \left(\mbox{hard} + \mbox{soft}\right) \, .$$
Therefore, a pixel in a HR image have values bounded in the interval $(-1, +1)$, i.e., from minimum hardness to maximum hardness. For a plasma emitting through bremsstrahlung process (such as the intracluster gas), the hardness will be proportional to the temperature. In this sense, the hardness ratio map is a quick route to obtain an approximate temperature map of the intracluster gas \citep[as done by, e.g.,][]{Lagana2010,Andrade-Santos2013}.

\begin{figure*}
\includegraphics[width=0.8\textwidth]{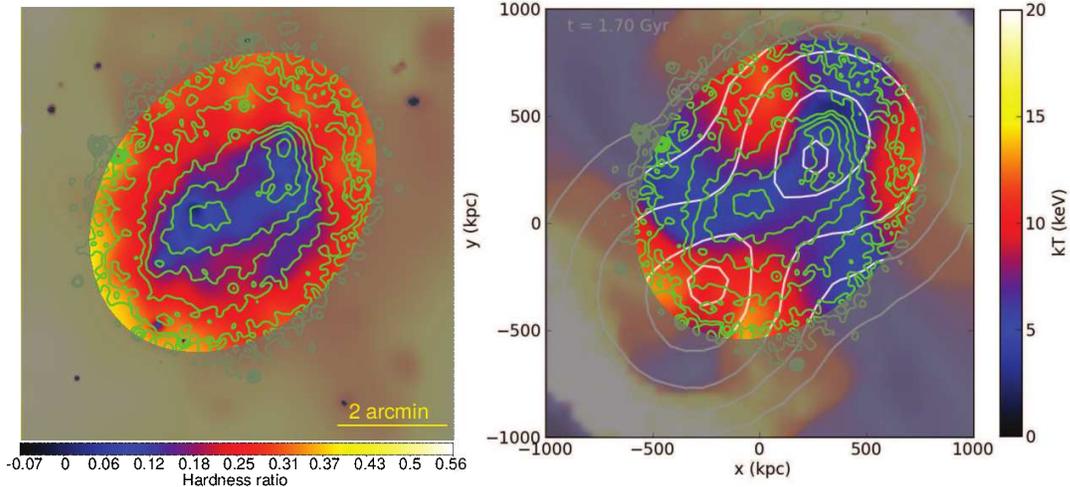}
\caption[]{\textsf{Left-panel}: Hardness ratio from \textit{Chandra} data (colors) and observed X-ray contours (green lines, the same as in Fig.~\ref{fig1}c). \textsf{Right-panel}:  Simulated temperature map (colors) with simulated contours of total mass (white lines) and observed X-ray contours (green lines). We have masked in both panels the region where the hardness ratio is dominated by the background.}
\label{fig6}
\end{figure*}

In Fig.~\ref{fig6}, left panel, we show the HR image of A1758N with a roughly similar colour table as the simulated temperature map shown on the right panel. At low count rate, the HR to temperature relation is not reliable since the data becomes dominated by the X-ray and particle background. Therefore, we have masked out the region with low, background corrected, count rate, 0.6 and 0.8 count/pixel$^2$ for the soft and hard band, respectively (1 pixel = 1.968~arcsec).

Comparing both panels of Fig.~\ref{fig6} we note a rough similarity, but also some differences. One cannot expect a pixel-by-pixel equivalency between simulated and observed data in such comparisons, let alone in the case of the temperature. Still, this procedure gives a qualitative idea of what would be seen, at least is some broad sense of hot regions versus cold regions. Roughly speaking, the peaks ok X-ray would be located in the inner colder part, while the warmer gas is seen in the outer parts of this field.

The fact remains that the predicted high temperatures associated with strong shocks are not observed in this cluster, even though several other features are adequately matched. This discrepancy could conceivably be due to shortcomings of the simulation, of the observation, or both. From the simulation side, there could be numerical reasons such as dependence on resolution or dependence on the hydrodynamics algorithm. However, increasing the number of particles does not weaken the shock. The limitations of the hydrodynamics algorithm, however, are less clear. SPH is known to have difficulty in capturing fluid instabilities under certain regimes \cite[see e.g.][]{Agertz2007}. Adaptive Mesh Refinement (AMR) is the other commonly used technique and it has its own difficulties, such as following small regions that move with high velocity \citep{Springel2010}, but it is known to offer better resolution at capturing shocks. In fact, \cite{Agertz2007} carried out systematic comparisons between SPH and AMR codes, using the test problem of a gas cloud moving through a gaseous medium. They show that in grid codes, the cloud tends to be more susceptible to fragmentation, whereas in SPH codes it tends to remain more cohesive. If the instabilities are not well captured in SPH, perhaps the insufficient mixing between the ICM gas of the two clusters could result in overestimated shocks. A thorough analysis of these issues is, however, beyond the scope of the present work. Even if the hydrodynamics algorithm and the resolution are both reliable, there could conceivably be some relevant physical processes in the real cluster that were not included in the simulation (magnetic fields, cooling, the effects of turbulence, non-thermal processes, etc). This type of binary collision is a highly idealized model, and the many simplifications were listed in Section \ref{sec:sim}. For example, real cluster mergers and cosmological simulations both have more complex flows due to the presence of substructure, which is absent in binary collisions. One of the difficulties of speculating on the lack of additional physics is that -- if there is some indispensable physical mechanism that is absent -- it would have to be some process capable of bringing substantial change on the largest scales.

Lastly, the inclination angle could possibly influence the projected temperatures. When the orbital plane of the merger is perpendicular to the line of sight, this gives the highest temperatures at the shocks. That is because the layers of shock-heated gas are seen as `edge-on' as possbile. Under an inclination, this hot gas is projected onto a larger area together with cooler gas from the surrounding regions. Inclination then decreases the amplitude of the temperature jumps, meaning that the Mach numbers estimated from observations are usually to be taken as lower limits. In the case of our model of A1758N, however, there was no reason to adopt an inclination. Attempts revealed that the morphology was generally degraded, and for the small inclination angles that would be tolerable, the decrease in temperature was not sufficient to justify it.


\subsection{Shock velocity}
\label{sec:shock}

In Section \ref{sec:global}, we saw that the relative velocity between the dark matter centroids is approximately 380 km s$^{-1}$ at $t = 1.7$ Gyr. Here we will see that the velocity of the shock wave is substantially higher at the same time. As explained by \cite{Springel2007} in the case of the Bullet Cluster, there is no obvious correspondence between the shock velocity and the velocity of the dark matter haloes. Indeed, as an extreme case of this behaviour, we find a very large shock velocity at an instant when the clusters are approaching the apocenter (i.e. vanishing relative velocity, along the radial coordinate). The shock wave propagates with a velocity that is different from the relative velocity between the two dark matter haloes, and furthermore this difference is time-dependent. In fact, as the collision progresses after the fist core passage, the dark matter haloes slow down, falling behind, while the shock fronts continue to propagate outwards at high velocities. In retrospect, these two velocities are conceptually distinct and there would be no justification to presume them equal at all times. By analyzing a dedicated hydrodynamical simulation, \cite{Springel2007} estimated a shock velocity of $\sim$4500 km s$^{-1}$ for the Bullet Cluster, nearly twice as fast as the motion of the mass centroids.

\begin{figure}
\includegraphics[width=\columnwidth]{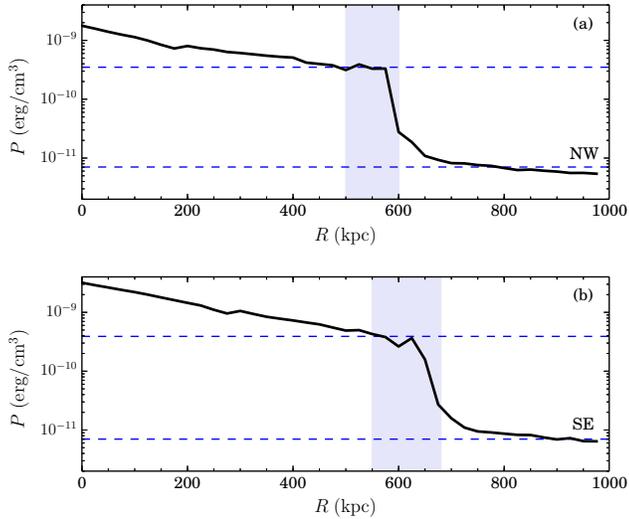}
\caption[]{Pressure profiles for the (a) NW and the (b) SE clusters, measured across the shock fronts (along the directions indicated by the arrows in Fig.~\ref{fig5}). The horizontal dashed lines show the amplitudes of the jumps at the discontinuity, as estimated from the dynamically measured Mach numbers.}
\label{fig7}
\end{figure}

In our model of A1758N we find important shock waves propagating outwards, ahead of each of the clusters. The very fact that the shock fronts are already much further away than the dark matter haloes is in itself an indication that the shock advances faster than the dark matter haloes. At the $t = 1.7$ Gyr snapshot, each shock front (the outer surfaces of the shocked gas; Fig.~\ref{fig5}) is almost 1 Mpc ahead of the respective halo centroid. From the simulation output, we have the necessary quantities to compute all relevant velocities. The following procedure is applyed independently to the NW shock front and to the SE shock front. First we measure temperature profiles along the directions that interceps the shock fronts as perpendicularly as possible (the black arrows in Fig.~\ref{fig5}). This is done at several time-steps before and after the $t = 1.7$ Gyr snapshot. The discontinuity in temperature is easily detectable, as the outer face of the shock, where temperature drops abruptly. Once the locations of these drops are known as a function of time, trivial numerical differentiation of the successive positions provides the velocities of the shock waves, repectively $v_{s} = 3054$ km s$^{-1}$ and $v_{s} = 3220$ km s$^{-1}$ for the NW and the SE fronts, at the relevant instant in time. However, following \cite{Springel2007}, we note that the pre-shock gas (i.e. the gas that has not yet been met by the shock front) is not at rest; rather, it falls towards the centre of mass. It is possible to measure the velocity of this upstream gas which is coming to meet the shock wave from the outside. We obtain, for NW and SE respectively, $u = -1440$ km s$^{-1}$ and $u = -1462$ km s$^{-1}$. Since $v_{s}$ is the velocity at which the shock front advances in the centre-of-mass rest-frame, and $u$ is the velocity of the upstrem gas in the centre-of-mass rest-frame, $v_{s}-u$ is the effective velocity with which the shock front encounters the pre-shock gas.

\begin{figure}
\centering
\includegraphics[width=0.8\columnwidth]{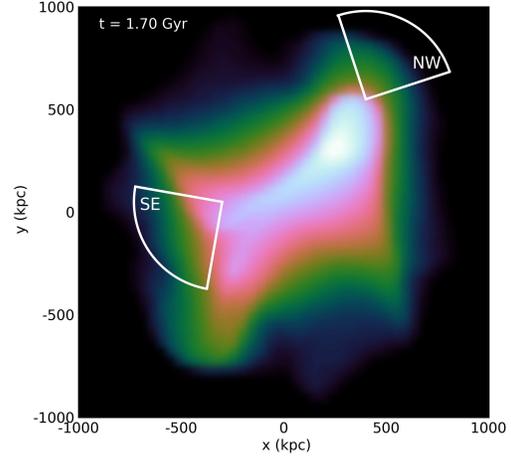}
\caption[]{Regions within which the pressure profiles of Fig.~\ref{fig9} were measured. Colors are the X-ray surface brightness.}
\label{fig8}
\end{figure}

\begin{figure}
\includegraphics[width=\columnwidth]{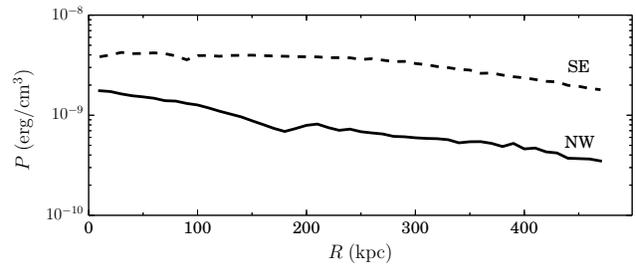}
\caption[]{Pressure profiles for the NW and the SE clusters, measured within the circular sectors displayed in Fig.~\ref{fig8}.}
\label{fig9}
\end{figure}

Once we know the effective shock velocities, the Mach numbers may be computed as:
\begin{equation}
\mathcal{M} =  \frac{v_{s}-u}{c_{s}}
\end{equation}
where $c_{s}^{2}=\frac{\gamma k T}{\mu m_{p}}$ is the sound speed of the pre-shock gas ($k$ is Boltzmann's constant, T is the temperature, $\mu$ is the mean molecular weight of the gas, and $m_{p}$ is the mass of the proton). The sound speed is thus obtained essentially by measuring the temperature of the upstream gas within a suitable region. Because the shock waves are advancing into the coldest parts of the ICM, the sound speed in those outer regions will be relatively low. We arrive at $c_{s} = 711$ km s$^{-1}$ and $c_{s} = 715$ km s$^{-1}$, for NW and SE respectively. From these, it follows that the Mach numbers are quite high, $\mathcal{M}=6.3$ and $\mathcal{M}=6.7$ for NW and SE respectively.

And yet the Mach numbers meausured according to this procedure are precisely consistent with the independently measured pressure jumps. Having measured the Mach numbers dynamically, we can use those result to compute the expected amplitudes of the pressure discontinuities at the locations of the shock fronts. The ratios between pre-shock (subscript 1) and post-shock (subscript 2) pressures are given by the corresponding Rankine-Hugoniot jump condition \citep[e.g.][]{Landau1959}:
\begin{equation}
\frac{P_{2}}{P_{1}} = \frac{5\mathcal{M}^{2}-1}{4} 
\end{equation}
where the adiabatic index $\gamma=5/3$ was assumed. Using the Mach numbers obtained above, we find that the pressures must drop by factors of approximately $P_{2}/P_{1} = 50$ and $P_{2}/P_{1} = 56$ for NW and SE. These amplitudes are plotted in Fig.~\ref{fig7}, where they clearly match the pressure profiles measured in the simulation (along the black arrows in Fig.~\ref{fig5}). The pressure profiles are measured within narrow cylinders that cross the shock front perpendicularly.

No such strong shocks are detected from the observational data, however. One should notice that the shocks in the simulation are located far away from the peaks of X-ray emission, in very low-density regions, where photon counts are expected to be extremely low (Fig.~\ref{fig1}c). From the observational point of view, it is conceivable that such features would be undetectable from the available data, for lack of a high-resolution temperature map covering a much larger area. \cite{David2004} reported no evidence of a shock front within 400 kpc of either subcluster in A1758N. In this regard, our simulated scenario is in full agreement: indeed the shocks are more than twice as far in the simulation and would not have been detected within 400 kpc. Our model also agrees with their conclusion that A1758N is in the late stages of the collision, as they correctly point out that the merger shocks have already passed through most of the cluster.

However, our simulation analysis departs from their interpretation of the strength of the shock, estimated as $\mathcal{M} \leq 1.15$. This is a salient point that we attempt to reconcile here. \cite{David2004} estimate a pressure jump by a factor of only $1.4 \pm 0.3$ across the NW edge and conclude that the relative velocity between the two subclusters should be less than about 1600 km s$^{-1}$. That the velocity of the subclusters cannot be constrained in this way, has already been made clear by \cite{Springel2007} and by our analysis. As far as the interpretation of the pressure jump is concerned, we offer a possibile argument that may connect the simulation output with what is available to be measured in the observational data.

If, in the simulation, we were to measure pressure profiles within the circular sectors marked in Fig.~\ref{fig8} -- rather than at the actual location of the shocks --, we would obtain the results shown in Fig.~\ref{fig9}. This exercise cannot be expected to accurately emulate the observational data points. Nevertheless, the smoothness of the curves in Fig.~\ref{fig9} may be used to argue in favor of the following interpretation: that it is conceivable that one might obtain pressure decrements of order unity if the regions used for measurements are immediately ahead of the X-ray peaks.

According to our model, the current dynamical stage of A1758N is such that the shock fronts are far removed from the dense inner regions. But this configuration does not need to hold true for any cluster merger. In fact, the Bullet Cluster itself is a good example of a merging cluster where deep X-ray observations do reveal a strong shock shortly after core passage. In their SPH simulation of the Bullet Cluster, \cite{Springel2007}, for example, measured a temperature jump from $\sim$30 keV to $\sim$10 keV at the bow shock position. This was in good agreement with the projected temperature profile measured by \cite{Markevitch2006}. That may be reagarded as a `strong' shock, at least in the sense that the temperature of the shock-heated gas is substantially high. Yet, the resulting Mach number of $\mathcal{M}\sim3$ is not exceedingly large because the pre-shock gas is also quite warm, meaning that the temperature jump is not too large (and neither is the sound speed). However, in that model the best instant takes place 0.18 Gyr after core passage, a time when the bow shock is only $\sim$70 kpc ahead of the bullet centroid, i.e. still within a high-density region, where X-ray emission is relatively abundant. Perhaps a similar situation should be expected for the El Gordo cluster, which is generally regarded as a high-redshift analog of the Bullet Cluster. Some simulations of the El Gordo merger have been carried out \citep{Donnert2014, Molnar2015}, but no detailed study of its shock and Mach number has been performed yet.

The absolute ages (or mass centroid separations) of different cluster mergers should not be directly compared, because the time scales depend on the dynamics of each system. Nevertheless, one could expect in general that the detectability of shock fronts must ultimately depend on the emissivity (i.e. essentially gas density) of the region where the bow shock is currently located. Then, there might be three reasons why such detections are relatively rare. First, a massive cluster is needed to obtain a strong shock. Second, it must be observed shortly after core passage to ensure sufficient X-ray emission. Third, the orbital plane should be close to the plane of the sky not to weaken the temperature jump in projection. Each of these requirements is separately unlikely: massive clusters are less abundant than low-mass ones; core passage is the moment of highest velocity, thus the time spent near the pericenter is short; and all orientations are possible, although a range of small inclinations is tolerable. According to our model, A1758N would seem to satisfy only the first and the third requirements.


\section{Discussion and summary}
\label{sec:summary}

We have offered one possible scenario for the merging event of A1758N. In our best model, two massive galaxy clusters -- of $\sim 5 \times 10^{14} M_{\odot}$ each -- undergo an off-axis collision. Our model recovers the general morphological feature that in one of the clusters the X-ray and mass peaks coincide, while in the other gas and dark matter are offset. This model accounts for several observed features, namely the current separation between the mass peaks, the global X-ray morphology, central gas densities and central gas temperatures. This scenario also entails the existence of two prominent bow shocks, which are not detected from currently available observations. It also introduces an apparently high shock velocity. Nevertheless we argue in favor of the plausibility of such a velocity, given the parameters of the merger. If this scenario can be unsdestood to be a fair representation of A1758N's history, then one must attempt to reconcile its predictions with the observations.

Starting from an initial separation of 3 Mpc, with impact parameter $b = 250$ kpc and relative velocity $v_{0} = -1500$ km s$^{-1}$, the clusters pass by each other at a pericentric distance of only about 60 kpc. Although the off-axis nature of the encounter might appear inexpressive, as suggested by such a small pericentric distance, in reality the angular momentum of the system -- set by the initial conditions -- is quite sufficient to bring about the necessary degree of asymmetry in the final configuration. The small pericentric distance would also suggest an important deflection angle of the trajectories in the case of point masses, but given the extended nature of the simulated objects, the deflection is in fact as small as $15^{\circ}$. One should also note that the $v_{0}$ employed is somewhat lower than the velocity of the zero-energy orbit. This means that this collision can hardly be said to be particularly violent, at least in terms of the relative velocity of the clusters. Approximately 0.4 Gyr after the first central passage, the clusters reach the required separation of 750 kpc, by which time they have decelerated almost to the returning point. Their relative velocity is then 380 km s$^{-1}$ and they are slowing down to approach the apocenter.

The morphology of the ICM is the most challenging feature to reproduce, specially while simultaneously satisfying all the other observational constraints. In our model, the broad curved shape of the X-ray emission is recovered, as well as the necessary offset between gas and dark matter in one of the clusters. The secondary X-ray peak does not appear as distinctly intense in the visualizations, but quantitatively, the measured gas densities in the simulation agree well with the values from \cite{David2004}. Our preferred model has no inclination with respect to the line of sight, i.e. the plane of the orbit coincides with the plane of the sky. This is so for the simple reason that no relevant morphological improvement was obtained by projecting the simulation under different angles. Nevertheless, small inclinations of $i \lesssim 20^{\circ}$ could produce similarly acceptable results. This would accomodate a line-of-sight relative velocity of up to $\sim$130 km s$^{-1}$.

The inner temperatures of approximately $T \sim 8$ keV for both clusters are well reproduced in or model. This is consistent with the masses of the clusters and with the fact that the gas densities were also in good agreement. However, the simulation implies the existence of shock-heated gas at high temperatures expanding towards the periphery of the cluster. This is not seen in the observations, however. In the simulation each shock wave has already propagated outward to very large distances, being found at nearly 1 Mpc away from the respective cluster centre. We argue that it is plausible that the high temperatures would be undetectable in the in those remote low-emission regions. Furthermore, it is not clear how one would produce a model where two clusters of $\sim 5 \times 10^{14} M_{\odot}$ and $\sim$8 keV could collide without giving rise to shock-heated gas. We cannot cover the parameter space of collisions broadly enough to strictly rule out alternative models. But given the reasonable orders of magnitude involved in this problem, we tend to favour a picture in which strong shocks did occur, but have already propagate to regions where their detectability is made difficult.

We have measured shock velocities of $\sim$4500 km s$^{-1}$ in the simulation. As we have pointed out, following \cite{Springel2007}, the velocity of the shock and the velocity of the clusters are entirely separate concepts, and may not be identified. In fact, we present an extreme case, in which the clusters themselves are nearly coming to a halt, while the shock front continues to advance with a substantial velocity. This illustrates again the roles played by numerical simulations of cluster mergers: not merely to model observed phenomena, but also to supply the means by which to correctly interpret observational results. We measured the Mach numbers dynamically, taking into account the infalling velocity of the upstream gas, and obtained $\mathcal{M}=6.3$ and $\mathcal{M}=6.7$. These are consistent with the amplitudes of the pressure jumps at the location of the shock front, corresponding to factors of about $\sim$50 given by the Rankine-Hugoniot conditions. No such thing is seen in the observations. But to detect such pressure discontinuities, one would have to be able to estimate density and temperature across the region of the shock fronts, which are not detectable. In an attempt to reconcile the simulations model with what is actually available to be measured, we showed what would be the outcome of measuring pressures in the region immediately ahead of the X-ray peaks. The results show that the decrements would be of order unity, rather than $\sim$50. Therefore, we argue in favor of the scenario where the shock fronts are very removed from the central regions, suggesting that if pressure discontinuities are sought in the vicinity of the X-ray peaks, it is not the actual shock front that will be found. From statistical studies of shock waves in cosmological simulations \citep[e.g.][]{Hoeft2008, Skillman2008}, it is found that larger shock waves are more often found in low-density regions and tend to have larger Mach numbers. Conversely, smaller shock waves occur nearer the cluster centres, having lower Mach numbers. This trend, by the way, is consistent with the rarity of central radio relics \citep{vanWeeren2009, Skillman2011, Vazza2012}. Given the specific circumstance of A1758N, our best model happened to be at a time when the shock fronts are already in distant regions, but in our simulations we generally see shocks passing through the cluster cores. However, their presence near the cores is short-lived, compared to the other phases of evolution.

We have offered one plausible scenario that is physically well-motivated and that satisfies a number of specific observational constraints. This scenario predicts properties of the shock front that cannot be detected with currently available data. It is of course impossible to argue for the uniqueness of the solution we have proposed. One cannot rule out the possibility that alternative combinations of collision parameters might provide comparable or even better results.


From the simulation standpoint, one prospect would be to rerun these simulations with an AMR code, since that scheme is known to capture fluid instabilities more accurately. The results could then either corroborate the prediction of strong shocks in this model of A1758N, or else they could reveal weaker shocks, thereby suggesting the inadequacy of the SPH approach in modelling certain aspects of this kind of cluster merger. From the observational side, the issue of the detectability of strong shocks in the low-density periphery of the cluster remains challenging since we would need roughly between 800~ks to 1~Ms exposure to achieve the necessary signal-to-noise ratio at the shock region in order to map the temperature through the surface brightness hardeness ratio.

With the simulation model presented here, we have shown that it is indeed possible to reproduce the very specific morphological features of A1758N by means of a major merger. In a forthcoming paper, we will provide improved estimates of the masses of each subcluster, derived from new gravitational weak lensing analysis. The new mass determinations will help fine-tune our dynamical model.

\section*{Acknowledgements}

This work has made use of the computing facilities of the Laboratory of Astroinformatics (IAG/USP, NAT/Unicsul), whose purchase was made possible by the Brazilian agency FAPESP (grant 2009/54006-4) and the INCT-A. REGM acknowledges support from FAPESP (2010/12277-9) and from \textit{Ci\^encia sem Fronteiras}. RMO acknowledges support from CAPES and CNPq. GBLM acknowledges support from CNPq. ESC acknowledges support from CNPq and FAPESP (2014/13723-3). We thank the referee for providing helpful suggestions that improved the paper.

\bibliographystyle{mn2e.bst}
\bibliography{a1758.bib}
\bsp
\label{lastpage}

\end{document}